\documentclass[aps,prc,reprint,amsmath,amssymb,nofootinbib,preprintnumbers]{revtex4-1}
\usepackage{graphicx}

\DeclareMathOperator{\Tr}{Tr}

\begin{document}
\preprint{BONN--TH--2014--17}
\title{Non-Geometric F-Theory--Heterotic Duality}
\author{Jie Gu}
\email{jiegu@th.physik.uni-bonn.de}
\author{Hans Jockers}
\email{jockers@uni-bonn.de}
\affiliation{
Bethe Center for Theoretical Physics\\
Physikalisches Institut, Universit\"at Bonn\\
53115 Bonn, Germany
}
\begin{abstract}
In this work we study the duality between F-theory and the heterotic string beyond the stable degeneration limit in F-theory and large fiber limit in the heterotic theory. Building upon a recent proposal by Clingher--Doran and Malmendier--Morrison --- which phrases the duality on the heterotic side for a particular class of models in terms of (fibered) genus two curves as non-geometric heterotic compactifications --- we establish the precise limit to the semi-classical heterotic string in both eight and lower space-time dimensions. In particular for six dimensional theories, we argue that this class of non-geometric heterotic compactifications capture $\alpha'$~quantum corrections to the semi-classical heterotic supergravity compactifications on elliptically fibered $K3$ surfaces. From the non-geometric heterotic theory, the semi-classical phase on the $K3$~surface is recovered from a remarkable limit of genus two Siegel modular forms combined with a geometric surgery operation. Finally, in four dimensions we analyze another limit deep in the quantum regime of the non-geometric heterotic string, which we refer to as the heterotic Sen limit. In this limit we can explicitly argue that the semi-classical two-staged fibrational structure of the heterotic hypermultiplet moduli space --- recently established by Alexandrov, Louis, Pioline and Valandro --- gets corrected by quantum effects.
\end{abstract}
\maketitle


\section{Introduction}
F-theory--heterotic duality remarkably relates in eight space-time dimensions F-theory on an elliptically fibered $K3$ surface to the heterotic string on the two torus \cite{Vafa:1996xn,Friedman:1997yq,Aspinwall:1997ye,MR2126482}. Upon adiabatically fibering this duality over suitable base spaces, the F-theory--heterotic duality generalizes to compactifications in lower space-time dimensions as well.\footnote{In particular, compactifying the eight-dimensional duality on an additional two torus, F-theory--heterotic duality relates to the type~IIA--heterotic duality in six space-time dimensions \cite{Witten:1995ex}.}

As we do not have a microscopic description of F-theory, the duality to the heterotic string has given us valuable insights into the physics of F-theory compactifications. The relationship of non-Abelian gauge groups in F-theory to the heterotic string via the spectral cover construction \cite{Friedman:1997yq,Donagi:2008ca} has led to a powerful toolbox to geometrically engineer (minimal) supersymmetric gauge theories \cite{Beasley:2008dc}.\footnote{For a review on this topic, see for instance ref.~\cite{Weigand:2010wm}.} As a result the dictionary between the F-theory compactification space and the low energy effective gauge theory particle spectrum is rather well understood. Furthermore, on the heterotic side the duality has shed light on certain non-perturbative aspects of the heterotic string --- such as NS5-branes states \cite{Witten:1995gx} --- that enjoy a geometric description in the dual F-theory description.

While F-theory--heterotic duality dictionary is fairly well established for (indexes of) particle spectra \cite{Beasley:2008dc}, it has been less explored on the level of moduli spaces for (quantum-exact) effective interactions. This is due to the fact that the F-theory--heterotic duality is often formulated in a certain limit  --- namely in a stable degeneration limit of the F-theory geometry and a large fiber limit of the heterotic compactification space \cite{Friedman:1997yq,Aspinwall:1997ye,Aspinwall:1998bw} --- which describes the duality only at the boundary of the moduli spaces of the dual theories.\footnote{See, for instance, refs.~\cite{LopesCardoso:1996hq,Lerche:1998nx,Berglund:1998ej,MR2126482,Berglund:2005dm,Jockers:2009ti}, where particular interactions are described beyond this limit.}

Building on earlier work \cite{LopesCardoso:1996hq,Lerche:1998nx,McOrist:2010jw} and by exploiting the Shioda--Inose structure of certain elliptically-fibered $K3$ surfaces \cite{MR2369941,MR2355598,MR2427457,MR2824841,MR2935386} --- based on the mathematical program pursued by Clingher and Doran --- Malmendier and Morrison \cite{Malmendier:2014uka} achieve to identify the moduli spaces (and thus the effective interactions) of a particular simple dual F-theory--heterotic pair in eight space-time dimensions, which describes the partial higgsing of the gauge group $E_8\times E_8$ to $E_7\times E_8$ of the associated low energy effective eight-dimensional supergravity theory. Resulting from the duality they find that the quantum-exact effective heterotic description is geometrically captured by a genus two curve and the low energy effective action is encoded in genus two Siegel modular forms \cite{Mayr:1995rx}. Upon further compatifying to lower space-time dimensions, it is further argued that such simple geometric F-theory scenarios give rise to non-geometric compactifications of the heterotic string \cite{McOrist:2010jw,Malmendier:2014uka}. These heterotic string theories nevertheless pose an effective geometric description in terms of genus-two fibered projective varieties, which are not Calabi--Yau varieties.

The aim of this article is to further analyze the Clingher--Doran--Malmendier--Morrison F-theory--heterotic correspondence \cite{Malmendier:2014uka}. We explicitly establish in this picture --- both in eight and in lower space-time dimensions  --- the limit to the semi-stable degeneration geometry on the F-theory side and the large fiber limit on the heterotic side. Establishing this semi-classical limit to the F-theory--heterotic duality provides for a natural starting point to study heterotic quantum corrections to its semi-classical large fiber approximation, as described by the non-geometric effective heterotic compactification space.

Deep in the non-geometric heterotic regime, we discover in the moduli spaces of genus two fibered effective heterotic compactifications the loci of constant (but non-trivial) genus two fibration. In analogy to the Sen limit in F-theory \cite{Sen:1997gv}, we call such loci the heterotic Sen limit. In particular, for non-geometric heterotic compactifications to six and four space-time dimensions, we connect the heterotic Sen limit to the structure of the hypermultiplet moduli space, as recently discussed in the semi-classical approximation in refs.~\cite{Louis:2011aa,Alexandrov:2014jua}. By comparing with the heterotic Sen limit, we give evidence that their proposed hierarchical fibrational structure in the hypermultiplet moduli space is a property of the semi-classical approximation in the absence of quantum corrections.

The organization of this article is as follows. In Section~\ref{sec:8d} we discuss the F-theory--heterotic correspondence in eight space-time dimensions. In particular, we establish the limit to the semi-classical approximation of the F-theory--heterotic duality in the F-theory stable degeneration limit and the heterotic large volume limit in terms of the Siegel operator acting on genus two Siegel modular forms.  In Section~\ref{sec:6Dduals} we consider the F-theory--heterotic correspondence in the non-geometric phase adiabatically fibered over a $\mathbb{P}^1$ base. Again we establish the limit to the F-theory--heterotic duality in the stable degeneration and the large fiber limit, where we recover the semi-classical heterotic geometric compactification on the elliptically-fibered $K3$ surface. Deep in the heterotic quantum regime, we define the heterotic Sen limit, which allows us to compare with the semi-classical two-staged fibrational structure of the hypermultiplet moduli space of four-dimensional heterotic compactifications studied by Alexandrov, Louis, Pioline and Valandro. In Section~\ref{sec:con} we present our conclusions.

\section{8d N=2 F-theory--heterotic duality} \label{sec:8d}
It was proposed in refs.~\cite{Vafa:1996xn,Friedman:1997yq,Aspinwall:1997ye} that F-theory compactified on an elliptic $K3$ surface with a section is dual to the $E_8\times E_8$ heterotic string theory compactifed on a torus $T^2$. Since then much work has been devoted to understand this duality. The moduli space of the heterotic string has complex dimension 18, combining the complex structure modulus $\tau$ and the complexified K\"{a}hler modulus $\rho = B+i J$ of the torus as well as the 16 complex Wilson lines. In the F-theory one can write down the Weierstrass model for the elliptic $K3$ surface with a section
\begin{equation}
		y^2 = x^3 +f_8(t)x+f_{12}(t)
\end{equation}
where $f_8(t)$ and $f_{12}(t)$ are polynomials of degrees $\leqslant 8$ and $\leqslant 12$ respectively of the affine coordiante $t$ on the $\mathbb{P}^1$ base. These are 22 parameters. After modding out the $SL(2,\mathbb{C})$ diffeomorphism on the base $\mathbb{P}^1$ as well as one overall rescaling, one finds 18 complex moduli in the F-theory as well. 

Beyond the simple match of dimensionalities, an exact dictionary between the moduli spaces of the two theories is only known in the large volume limit on the heterotic side, which corresponds to the stable degeneration limit on the F-theory side \cite{Friedman:1997yq,Aspinwall:1997ye,Aspinwall:1998bw}.  In ref.~\cite{LopesCardoso:1996hq} a dictionary was given when all the Wilson lines are turned off. Clingher--Doran and Malmendier--Morrison recently extended this result to one non-trivial Wilson line in the heterotic theory \cite{MR2824841,MR2935386,Malmendier:2014uka}. Our first task is to analyze their proposal in further detail by exhibiting its relation to the duality in the large volume/stable degeneration limit.

\subsection{Moduli space of the $E_8\times E_7$ heterotic string} \label{sec:mspace}
The $E_8\times E_8$ heterotic theory on $T^2$ with one non-trivial Wilson line has the unbroken gauge group $E_8\times E_7$. Its moduli space is parametrized by two complex moduli $\tau,\rho$ and the complex Wilson line $z$, and it is given by \cite{Narain:1985jj,Malmendier:2014uka} \footnote{Here we follow ref.~\cite{Malmendier:2014uka} and use the index 2 subgroup $O^+(\Lambda^{2,3})$ of the full duality group $O(\Lambda^{2,3})$, as the former is the maximal subgroup for which modular forms are holomorphic.}
\begin{equation} \label{eq:ModHet}
		\mathcal{M}_{\textrm{het}} = \mathcal{D}_{2,3}/O^+(L^{2,3}) \ .
\end{equation}
with the Teichm\"uller space
\begin{equation}
		\mathcal{D}_{2,3} = (O(2)\times O(3)) \backslash O(2,3) \ .
\end{equation}
The action of the U-duality group~$O^+(L^{2,3})$ mixes all moduli $\tau,\rho, z$, where the U-duality lattice $L^{2,3}$ of signature $(2,3)$ is the orthogonal complement of the sublattice $E_8(-1)\oplus E_7(-1)$ in the unique even unimodular lattice $\Lambda^{2,18}$ of signature $(2,18)$. The sublattice $E_8(-1)\oplus E_7(-1)$ is associated to the remaining 17 Wilson lines to be set to zero.

The Teichm\"uller space $\mathcal{D}_{2,3}$ is isomorphic to the genus two Siegel upper half-space \cite{MR3235787}
\begin{equation}
    \mathbb{H}_2 \,=\,\left\{\underline{\tau} = \begin{pmatrix} \tau & z \\ z & \rho \end{pmatrix} \middle | \,\operatorname{Im}(\underline{\tau})\ \text{pos. definite} \right\} \ ,
\end{equation}
where $\tau,\rho,z\in \mathbb{C}$. The duality asserts that the Siegel threefold $\mathcal{A}_2 = \mathbb{H}_2/Sp(4,\mathbb{Z})$, which is a partial compactification of the moduli space of genus two curves, is isomorphic to the moduli space $\mathcal{M}_{\textrm{het}}$.

The Siegel threefold $\mathcal{A}_2$ has a boundary $H_0$ and two singular surfaces of $\mathbb{Z}_2$-orbifold singularities known as the Humbert surfaces $H_1$ and $H_4$. We refer to them as divisors $H_0$, $H_1$ and $H_4$, respectively, in both the compactified moduli space $\overline{\mathcal{A}}_2$ of genus two curves and the compactified heterotic moduli space $\overline{\mathcal{M}}_{\textrm{het}}$. The boundary $H_0$ at $\rho\to i \infty$ is associated to a singular genus two curve with a nodal point. At the Humbert surface $H_1$ at $z=0$ the genus two curve degenerates into two genus one components transversely intersecting in a single double point, whereas the Humbert surface $H_4$ at $\tau=\rho$ describes a (generically) smooth genus two curve with a non-generic $\mathbb{Z}_2$-automorphism (in addition to the hyperelliptic involution). In the heterotic theory, the divisor $H_0$ maps to the large volume limit, at the surface $H_1$ the Wilson line is turned off, and at the surface $H_4$ a $\mathbb{Z}_2$-quantum symmetry emerges.

As the Siegel threefold $\mathcal{A}_2$ is parametrized by the ring of genus two Siegel modular forms of even weight generated by the modular forms $\psi_4, \psi_6, \chi_{10}, \chi_{12}$ with weights $4,6,10,12$, respectively \cite{MR0141643}, it is tempting to identify the heterotic moduli space $\overline{\mathcal{M}}_{\textrm{het}}$ with the weighted projective space $\mathbb{P}^3_{(2,3,5,6)}$ of homogeneous coordinates $\psi_4, \psi_6, \chi_{10}, \chi_{12}$. We will illustrate momentarily that this description is not quite complete.

Let us recall the definition of the modular forms $\psi_{2k}$ of the classical Siegel Eisenstein series of weight $2k$
\begin{equation}
		\psi_{2k}(\underline{\tau}) = \sum_{(C,D)}\det(C \underline{\tau}+D)^{-2k}, \quad k\in\mathbb{Z_+} \ .
\end{equation}
Here the sum is taken over all equivalence classes of pairs $(C,D)$ of symmetric $2\times 2$ integral matrices subject to the equivalence relation $(C, D) \sim (M\cdot C, M\cdot D)$ in terms of any $SL(2,\mathbb{Z})$-matrix $M$. Furthermore, $\chi_{10}, \chi_{12}$ are two Siegel cusp forms defined by \cite{MR2772480,MR2935386}
\begin{equation} \label{eq:Defchi}
\begin{aligned}
		\chi_{10}&=\frac{43867}{2^{12} \, 3^5 \, 5^27 \cdot 53} (\psi_4\cdot \psi_6 - \psi_{10}) \ , \\
		\chi_{12}&=\frac{131\cdot 593}{2^{13} \, 3^7 \, 5^3 \, 7^2 \, 337} (3^2 \, 7^2 \psi_4^3\!+\!2\cdot 5^3 \psi_6^2\! -\! 691 \psi_{12}) \, ,
\end{aligned}
\end{equation}
which --- as Siegel cusp forms --- are mapped to zero under the Siegel operator
\begin{equation} \label{eq:SOcusp}
		(\Phi\chi_k)(\underline{\tau}) = \lim_{\rho \rightarrow i\infty} \chi_k \begin{pmatrix} \tau & 0 \\ 0 & \rho \end{pmatrix}  = 0 \ .
\end{equation}
The ring of all genus two Siegel modular forms has one more generator namely a cusp form $\chi_{35}$ of odd weight \cite{MR0229643}.\footnote{The ring of all genus two Siegel modular forms describes a double cover of the Siegel threefold $\mathcal{A}_2$ as the modular forms of odd degree reverse their sign with respect to the modular transformation of the matrix $\operatorname{Diag}(+1,-1,+1,-1) \in Sp(4,\mathbb{Z})$. This implies that a given genus two curve uniquely specifies the value of all even modular forms, but determines the odd modular forms only up to a sign.}
It is related to the Siegel modular forms of even weight by \cite{MR0229643}
\begin{equation}
\begin{aligned}
		\chi_{35}^2 &= \frac{1}{2^{12} \, 3^9} \,\chi_{10}\left(  2^{24}\,3^{15} \,\chi_{12}^5 -2^{13}\,3^9\,\psi_4^3\chi_{12}^4-2^{13}\,3^9\,\psi_6^2\,\chi_{12}^4  \right.\\
		&+3^3\,\psi_4^6\,\chi_{12}^3-2\cdot 3^3\, \psi_4^3\,\psi_6^2\,\chi_{12}^3- 2^{14}\,3^8\,\psi_4^2\,\psi_6\,\chi_{10}\,\chi_{12}^3 \\
		&-2^{23}\,3^{12}\,5^2\,\psi_4\,\chi_{10}^2\,\chi_{12}^3+3^3\,\psi_6^4\,\chi_{12}^3+2^{11}\,3^6\,37\,\psi_4^4\,\chi_{10}^2\,\chi_{12}^2 \\
		&+2^{11}\,3^6\,5\cdot 7\,\psi_4\,\psi_6^2\,\chi_{10}^2\,\chi_{12}^2-2^{23}\,3^9\,5^3\,\psi_6\,\chi_{10}^3\chi_{12}^2 \\
		&-3^2\psi_4^7\,\,\chi_{10}^2\,\chi_{12}+2\cdot 3^2\,\psi_4^4\,\chi_6^2\,\chi_{10}^2\,\chi_{12} \\
		&+2^{11}\,3^5\,5\cdot 19\,\psi_4^3\,\psi_6\,\chi_{10}^3\,\chi_{12} +2^{20}\,3^8\,5^3\,11\,\psi_4^2\,\chi_{10}^4\,\chi_{12}\\
		&- 3^2\,\psi_4\,\psi_6^4\,\chi_{10}^2\,\chi_{12} + 2^{11}\,3^5\,5^2\,\psi_6^3\,\chi_{10}^3\,\chi_{12} - 2\,\psi_4^6\,\psi_6\,\chi_{10}^3 \\
		& - 2^{12}\,3^4\,\psi_4^5\,\chi_{10}^4+2^2\,\psi_4^3\,\psi_6^3\,\chi_{10}^3 + 2^{12}\,3^4\,5^2\,\psi_4^2\,\psi_6^2\,\chi_{10}^4\\
		&\left.+2^{21}\,3^7\,5^4\,\psi_4\,\psi_6\,\chi_{10}^5-2\,\psi_6^5\chi_{10}^3 +2^{32}\,3^9\,5^5\,\chi_{10}^6 \right)\ .
\end{aligned}
\end{equation}

To further study our moduli spaces, we now locate the three divisors $H_0,H_1,H_4$ in the space $\mathbb{P}^3_{(2,3,5,6)}$. The surface $H_1$ at $z=0$ is given by $\chi_{10}=0$, while the surface $H_4$ at $\tau=\rho$ becomes $\frac{\chi_{35}^2}{\chi_{10}}=0$, which is a polynomial constraint in the (even) generators $\psi_4,\psi_6,\chi_{10},\chi_{12}$.

The large volume divisor $H_0$ is more complicated. The Siegel operator sends a Siegel cusp form to zero (c.f., eq.~\eqref{eq:SOcusp}) and maps a Siegel Eisenstein series to an elliptic Eisenstein series of the same weight
\begin{equation} \label{eq:SOEisen}
  (\Phi\psi_{2k})(\underline{\tau})
  = \lim_{\rho \rightarrow i\infty} \psi_{2k} \begin{pmatrix} \tau & 0 \\ 0 & \rho \end{pmatrix}
  = E_{2k}(\tau) \ .
\end{equation}
Furthermore, due to the periodicity of $\rho$, a Siegel modular form $\phi$ of weight $k$ enjoys the Jacobi--Fourier development
\begin{equation}
		\phi_k = \sum_{m=0}^{\infty} \phi_{k,m}(\tau,z) e^{2\pi i m \rho} \ .
\end{equation}
$\phi_{k,m}(\tau,z)$ is a Jacobi form of weight $k$ and index $m$, which satisfies the two modular transformation properties
\begin{equation}
\begin{aligned}
  &\phi_{k,m}(\tfrac{a\tau+b}{c\tau+d},\tfrac{z}{c\tau+d}) = (c\tau+d)^k e^{\frac{2\pi i m c z^2}{c\tau+d}}\phi_{k,m}(\tau,z)	\ , \\
  &\phi_{k,m}(\tau,z+\lambda\tau+\mu) = e^{-2\pi i m(\lambda^2\tau+2\lambda z)}\phi_{k,m}(\tau,z) \ , 
\end{aligned}
\end{equation}
with $\lambda,\mu\in \mathbb{Z}$ and $\begin{pmatrix} a & b \\ c & d \end{pmatrix} \in SL(2,\mathbb{Z})$. Since $\phi_{k,0}(\tau,z)$ of index 0 is independent of $z$ (it is an elliptic modular form of weight $k$), we arrive with eqs.~\eqref{eq:SOcusp} and \eqref{eq:SOEisen} at
\begin{equation} \label{eq:psilim}
  \lim_{\rho\rightarrow i\infty}\psi_{2k}(\underline{\tau}) = E_{2k}(\tau) \ , 
  \quad \lim_{\rho\rightarrow i\infty}\chi_{k}(\underline{\tau}) = 0 \ .
\end{equation}
The coefficients in the full Jacobi--Fourier development of the Siegel Einstein series are given in ref.~\cite{Eichler:1985zg}
\begin{equation}
		E_{2k}(\underline{\tau}) = \sum_N a(N)e^{2\pi i \Tr(N \underline{\tau})} \ . 
\end{equation}
The sum is taken over all half-integral symmetric matrices $N$ with integral entries on the diagonal. Combined with the definition \eqref{eq:Defchi}, we find
\begin{equation}
\begin{aligned}
  \lim_{\rho\rightarrow i\infty} \frac{\chi_{12}}{\chi_{10}} =& -\frac{\zeta ^2+10 \zeta +1}{3 (\zeta -1)^2} -\frac{4  (\zeta -1)^2 q}{\zeta }\\
  &-\frac{4 (\zeta\!+\!2) (2 \zeta \!+\!1) (\zeta\!-\!1)^2 q^2}{\zeta ^2} + \mathcal{O}(q^3) \ ,
\end{aligned}
\end{equation}
where $q=e^{2\pi i \tau}$ and $\zeta = e^{2\pi i z}$. This series agrees with the expansion of $\frac{\wp(z;\tau)}{\pi^2}$ in $q$ and $\zeta$, where $\wp(z;\tau)$ is the Weierstrass $\wp$-function. As $\lim_{\rho\rightarrow i\infty}\frac{\chi_{12}}{\chi_{10}}$ and $\frac{\wp(z;\tau)}{\pi^2}$ have the same modular properties, they must be identical, i.e., 
\begin{equation} \label{eq:chilimit}
		\lim_{\rho\rightarrow i\infty} \frac{\chi_{12}(\underline{\tau})}{\chi_{10}(\underline{\tau})} = \frac{\wp(z;\tau)}{ \pi^2 }  \ .
\end{equation}
Therefore, in the large volume limit the homogeneous coordinates $(\psi_4,\psi_6,\chi_{10},\chi_{12})$ of $\mathbb{P}^3_{(2,3,5,6)}$ become $(E_4,E_6,0,0)$, which furnishes a subvaritiety $h_0$  of codimension two. It is located within the non-transverse intersection of the divisors $\chi_{10}$ and $\frac{\chi_{35}^2}{\chi_{10}}$ of the Humbert surfaces $H_1$ and $H_4$. 
As $h_0$ has not the correct dimensionality to be identified with the boundary divisor~$H_0$, we consider the blow-up of $\mathbb{P}^3_{(2,3,5,6)}$ along the subvariety $h_0$
\begin{equation} \label{eq:blowup}
  \pi: \ \operatorname{Bl}_{h_0}\!\mathbb{P}^3_{(2,3,5,6)} \to \mathbb{P}^3_{(2,3,5,6)} \ ,
\end{equation}  
explicitly realized by the incidence correspondence
\begin{equation}
 \begin{aligned}   
    \operatorname{Bl}_{h_0}&\!\mathbb{P}^3_{(2,3,5,6)} \\
    =&\left\{ (I_0,I_2,\psi_4,\psi_6,\chi_{10},\chi_{12}) \, \middle|\, I_2 \chi_{10} - I_0 \chi_{12} = 0 \right\} \ .
\end{aligned}    
\end{equation}
Here $(I_0,I_2,\psi_4,\psi_6,\chi_{10},\chi_{12})$ denote the homogeneous coordiantes of the total space $\mathbb{P}(\mathcal{O}\oplus\mathcal{O}(1)) \to \mathbb{P}^3_{(2,3,5,6)}$, where the projective fiber coordinates $(I_0,I_2)$ have weight $(0,1)$ with respect to the base space $\mathbb{P}^3_{(2,3,5,6)}$.

Taking now the limit $\rho\to i\infty$ in the blown-up moduli space $\operatorname{Bl}_{h_0}\!\mathbb{P}^3_{(2,3,5,6)}$, we approach the fiber of the exceptional divisor $\pi^{-1}(h_0)$ with the coordinate ratio $\lim_{\rho\to i\infty}\frac{\chi_{12}}{\chi_{10}}=\frac{I_2}{I_0}=\frac{\wp(z;\tau)}{\pi^2}$. We claim that the exceptional divisor $\pi^{-1}(h_0)$ maps to the boundary divisor $H_0$, and that the blow-up space $\operatorname{Bl}_{h_0}\!\mathbb{P}^3_{(2,3,5,6)}$ describes the compactified heterotic string moduli space, i.e., 
\begin{equation} \label{eq:HMod}
   \operatorname{Bl}_{h_0}\!\mathbb{P}^3_{(2,3,5,6)} \simeq \overline{\mathcal{M}}_\text{het} \ .
\end{equation}   

While blowing-up the codimension two locus $h_0$ is a natural operation to arrive at the moduli space $\overline{\mathcal{M}}_\text{het}$, a detailed matching of moduli spaces would require a comparison to the compactification procedure of the Siegel threefold $\mathcal{A}_2$ as developed in refs.~\cite{MR0118775, MR0216035}. This is beyond the scope of this note. Instead, we will give a physics argument in the next subsection that justifies the identification of $\operatorname{Bl}_{h_0}\!\mathbb{P}^3_{(2,3,5,6)}$ with the moduli space $\overline{\mathcal{M}}_\text{het}$.
 
\subsection{Clingher--Doran--Malmendier--Morrison construction}
Using the Shioda--Inose structure \cite{MR2427457,MR2824841,MR2935386} on $K3$ surfaces with Picard lattices of rank $17$, in ref.~\cite{Malmendier:2014uka} Malmendier and Morrison determined the dictionary between the moduli space of F-theory compactifed on an elliptic $K3$ surface with $II^*$ and $III^*$ singular fibers and that of the $E_8\times E_8$ heterotic string compactified on $T^2$ with one non-trivial Wilson line. Starting from the Weierstrass model for the elliptically fibered $K3$ surface
\begin{equation}
		y^2 = x^3 + (a \, t^4 + b \, t^3)x +(c \, t^7+ d \, t^6+ e \, t^5) \ ,
\end{equation}
with affine coordinate $t$ on the base $\mathbb{P}^1$, the singular fibers $III^*$ and $II^*$ are located over $t=0$ and $t=\infty$, respectively. Then the five coefficients can be identified with the Siegel modular forms
\begin{equation}\label{equ:Malmendier}
\begin{aligned}
		a &= -\frac{1}{48} \psi_4(\underline{\tau})  \ , &
		b &= -4\chi_{10}(\underline{\tau}) \ , & 
		c &= 1 \ , \\
		d &= -\frac{1}{864}\psi_6(\underline{\tau}) \ , &
		e &= \chi_{12}(\underline{\tau}) \ .
\end{aligned}
\end{equation}
Together with a choice of symplectic basis, they uniquely fix the periods $\underline{\tau}$ of a genus two curve, which describe the moduli $(\tau,\rho,z)$ of the discussed heterotic string theory. Here, $\tau$ and $\rho$ are the complex structure and the volume modulus of $T^2$, whereas $z$ is the Wilson line modulus.

We want to trace this correspondence to the stable degeneration limit of the K3 surface, so as to argue that the blow-up~\eqref{eq:blowup} describes the physical moduli space $\overline{\mathcal{M}}_\text{het}$ of the dual heterotic string theory. In this limit the $\mathbb{P}^1$-base of the $K3$ surface splits into two $\mathbb{P}^1$s intersecting transversely at one point. Correspondingly, the elliptic $K3$ surfaces splits into two elliptic rational surfaces $S_1$ and $S_2$, which intersect at an elliptic curve $E$ \cite{Aspinwall:1997ye}. This elliptic curve $E$ becomes the two torus of the heterotic theory, while $S_1$ and $S_2$ determine the bundle data over $T^2$. To extract $E$ and $S_1, S_2$, we perform the change of variables
\begin{equation}
		(x, y, t) \mapsto \left(\frac{t^2 x}4, \frac{t^3 y}{16}, \pi t\right) \ ,
\end{equation}
and the Weierstrass model becomes \cite{Berglund:1998ej}
\begin{equation}
		t\, p_{+1} + p_0 + t^{-1} p_{-1} = 0 \ .
\end{equation}
where
\begin{equation}
\begin{aligned}
		&p_{+1} = 2^8 \pi^7, \\
		&p_0 = -y^2 + 4x^3 - \frac{4}{3}\pi^4 \psi_4(\underline{\tau})\, x - \frac{8}{27}\pi^6 \psi_6(\underline{\tau}) , \\
		&p_{-1} = -  2^8 \pi^3 \chi_{10}(\underline{\tau})\, x + 2^8 \pi^5 \chi_{12}(\underline{\tau}) \ .
\end{aligned}
\end{equation}
Here, $p_0$ is the elliptic curve~$E$, whereas $p_{\pm 1}$ encode the bundle data. In the discussed limit $p_0$ simplifies to
\begin{equation}
		y^2 = 4 x^3 - g_2(\tau) x - g_3 (\tau) \ ,
\end{equation}
with $g_2(\tau) = \frac{4\pi^4}3 E_4(\tau)$ and $g_3(\tau) = \frac{8\pi^6}{27} E_6(\tau)$, and describes an elliptic curve with period $\tau$. Furthermore, with eq.~\eqref{eq:chilimit} the intersection $E\cap\{ p_{-1}=0\}$ yields two points $(x,y) = (\wp(z;\tau),\pm\wp'(z;\tau))$ on the elliptic curve $E$ that add up to zero, which characterize the $SU(2)$ bundle over $E$ in the heterotic string \cite{Friedman:1997yq}, associated to the Wilson line modulus $z$.

On the heterotic side, the stable degeneration limit corresponds to the large volume limit $\rho \to i \infty$ of the two torus. Hence, we can compare this limit to our proposal~\eqref{eq:HMod} for the compactified heterotic moduli space. In the space $\operatorname{Bl}_{h_0}\!\mathbb{P}^3_{(2,3,5,6)}$, this limit maps to the exceptional divisor $H_0=\pi^{-1}(h_0)$, where the point in $h_0$ yields the complex structure of $E$, whereas --- up to a trivial rescaling --- the ratio of the coordinates $\frac{I_2}{I_0} = \frac{\wp(z;\tau)}{\pi^2}$ determines via the two points $(\wp(z;\tau),\pm\wp'(z;\tau))$ on $E$ the $SU(2)$ bundle data of the heterotic string.

This verifies that the proposed compactification \eqref{eq:HMod} of the moduli space $\overline{\mathcal{M}}_\text{het}$ agrees with the physical data in the large volume limit of the heterotic string. Furthermore, it demonstrates explicitly that the Clingher--Doran--Malmendier--Morrison construction \eqref{equ:Malmendier} is consistent with the F-theory--heterotic duality in the stable degeneration/large volume limit.

\section{6D N=1 F-Theory-Heterotic Duality} \label{sec:6Dduals}
In eight dimensions the Clingher--Doran--Malmendier--Morrison description encodes the complex structure modulus $\tau$, the K\"{a}hler modulus $\rho$ of the torus as well as the bundle modulus $z$ of the 8d heterotic theory in the moduli space of a genus two curve. To arrive at a 6d heterotic string theory, we can now adiabatically fiber this 8d construction over a suitable base. Hence, either we fiber the torus together with the bundle data over the base --- to obtain the heterotic string  on an elliptically fibered surface with a bundle --- or we directly fiber the three moduli $(\tau, \rho, a)$ over the same base to describe the heterotic string in terms of a genus two fibered surface. In the latter case the $Sp(4,\mathbb{Z})$ transition functions of the fiberation would mix up the three moduli in going from one coordinate patch to another. Therefore, there is no global distinction anymore between the complex structure, the K\"{a}hler and the bundle moduli. In other words, the complex structure moduli space of the genus two fibration combined with the K\"{a}hler moduli space of the base, constitutes the total moduli space of a \emph{non-geometric compactification} of the heterotic theory.

Let us fiber both sides of the Clingher--Doran--Malmendier--Morrison correspondence over a common $\mathbb{P}^1$ base. In the non-geometric description of the heterotic string --- given in terms of the genus two fibration $S\to \mathbb{P}^1$ --- the complex structure moduli of $S$ together with the volume of the base $\mathbb{P}^1$ are the heterotic moduli fields appearing in hypermultiplets. This non-geometric compactification is deep in the quantum regime of the heterotic string on a $K3$ surface with  finite volume, and $\alpha'$ corrections to the hypermultiplet moduli space play an important role. The dual F-theory description arises from an elliptically-fibered and K3-fibered Calabi--Yau threefold~$X$. Malmendier and Morrison show that the threefold~$X$ is an elliptic fibration over the Hirzebruch surface $\mathbb{F}_{12}$ \cite{Malmendier:2014uka}. This F-theory compactification of $X$ is actually familiar, as it appears in the conventional formulation of the F-theory--heterotic duality \cite{Morrison:1996pp}. This comes as a surprise, namely both phases of the heterotic string --- the geometric semi-classical and the non-geometric quantum regime --- are dual to the same semi-classical geometric F-theory compactification, and there is no appearance of a non-geometric F-theory phase.

To resolve this mystery, we consider the corresponding type IIA--heterotic duality \cite{Witten:1995ex}, i.e., type~IIA on the same Calabi--Yau threefold $X$ is dual to the geometric heterotic string on $K3\times T^2$ or --- as we claim --- to the non-geometric heterotic string on $S\times T^2$. The discussed 6d hypermultiplets become 4d hypermultiplets. In type~IIA on the threefold $X$, the hypermultiplet sector receives corrections in the string coupling $g_s$ \cite{Gunther:1998sc}. However, as we fiberwise apply the 8d type~IIA--heterotic duality in the adiabatic limit, the volume of the heterotic $\mathbb{P}^1$ is taken to be large. On the type~IIA side this amounts to working in the weak string coupling limit $g_s\to 0$ \cite{Witten:1995ex}. As a consequence, the tree level result of the type~IIA compactification is a good approximation for the hypermultiplets. In particular, it is legitimate to approximate the hypermultiplet sector metric $ds^{2,\textrm{tree}}_{\textrm{IIA HM}}$ by the classical c-map, which (in type~IIA) yields the tree level metric from the Weil--Petersson metric of the complex structure moduli space of $X$ \cite{Cecotti:1988qn,Ferrara:1989ik}. As the Weil--Petersson metric of the complex structure moduli space does not receive any $\alpha'$ corrections, we also do not expect any $\alpha'$ effects in the hypermultiplet sector --- at least so long as $g_s$-corrections are suppressed.

Therefore, in decompactifying the torus $T^2$, we expect that the absence of quantum corrections carries over to the 6d hypermultiplets of the F-theory compactification. In this way two heterotic phases of rather distinct nature enjoy a unifying semi-classical geometric F-theory description. Therefore, the discussed dual string theories furnish a remarkable example, where a non-trivial string duality gives rise to a conventional geometric description for a complicated dual non-geometric quantum phase.

\subsection{Genus two fibration} \label{sec:g2fib}
To arrive at the non-geometric heterotic theory in six dimensions, we start with the Weierstrass model on the F-theory side \cite{MR0229643,Malmendier:2014uka}
\begin{equation}\label{equ:6dWeierstrassModel}
\begin{aligned}  
  y^2 = x^3 - & \left(\frac{1}{48}\psi_4(s) t^4+4\chi_{10}(s) t^3\right)x  \\
  & \qquad + \left(t^7 -\frac{1}{864}\psi_6(s) t^6 + \chi_{12}(s) t^5\right)\ ,
\end{aligned}  
\end{equation}
with the affine coordinate $s$ on the $\mathbb{P}^1$ base. $\psi_4(s),\psi_6(s),\chi_{10}(s), \psi_{12}(s)$ are now sections of the line bundles $\mathcal{O}(8),\mathcal{O}(12),\mathcal{O}(20),\mathcal{O}(24)$ over $\mathbb{P}^1$, respectively. Note that the degrees of the line bundles are twice the weights of the associated modular forms.

The fibered modular forms $\psi_4(s),\psi_6(s),\chi_{10}(s), \psi_{12}(s)$ determine a genus two fibration over the same $\mathbb{P}^1$ base by assigning to the quadruple of modular forms the associated genus two curve. To arrive at an explicit description for this genus two fibered surface, we start with the hyperelliptic form of genus two curves
\begin{equation} \label{eq:hyper}
  y^2 =  x^6 + c_5(s) x^5 + \ldots + c_0(s) = \prod_{\ell=1}^6 \left(x- \xi_\ell(s)\right) \ .
\end{equation}  
Here the coefficients $c_\ell$ are elementary symmetric polynomials of the roots $\xi_\ell$, which are both fibered over the $\mathbb{P}^1$ base. Then the roots $\xi_\ell$ define the Igusa--Clebsch invariants $I_2$, $I_4$, $I_6$ and $I_{10}$ of respective weights  $2$, $4$, $6$, and $10$ \cite{MR0229643,MR0141643}\footnote{In the original references the numerical factors are absent in the definitions of $I_2, I_4, I_6$, because there the sums are taken over subsets of $S_6$ with indices $48, 72, 12$, respectively.}
\begin{equation}
\begin{aligned}
  I_2 &=\frac1{48} \sum_{\sigma\in S_6}(12)(34)(56) \ , \\
  I_4 &=\frac1{72} \sum_{\sigma\in S_6}(12)(23)(31)(45)(56)(64) \ , \\
  I_6 &=\frac1{12}\!\sum_{\sigma\in S_6}\!(12)(23)(31)(45)(56)(64)(14)(25)(36)\  , \\
  I_{10} &= \prod_{i<j} (\xi_i-\xi_j)^2 \ .
\end{aligned}   
\end{equation}
Here we define $(ab)\equiv (\xi_{\sigma(a)} - \xi_{\sigma(b)})^2$, and the sum is taken over all permutations $\sigma$ of the symmetric group $S_6$. Furthermore, the Igusa invariants relate to the Siegel modular forms according to \cite{MR0229643}
\begin{equation}
\begin{aligned}
  \psi_4 &= 2^4\cdot 3^2 I_4 \ , & \psi_6& =2^6 3^3(3 I_6 - I_2I_4) \ , \\
  \chi_{10}& = 2\cdot 3^5 I_{10} \ , \qquad & \chi_{12} &= 2\cdot 3^5 I_2 I_{10} \ . 
\end{aligned}  
\end{equation}
Thus the coefficients of the hyperelliptic equation~\eqref{eq:hyper} become functions of modular forms
\begin{equation}
  c_\ell = c_\ell(\psi_4,\psi_6,\chi_{10},\chi_{12}) \ , \quad \ell=0,\ldots,5 \ ,
\end{equation}
that are implicitly defined by their relationship to the roots $\xi_\ell$. In this way get a rather indirect description of the genus two fibered surface $S$. It would be interesting to have a more explicit description of the surface $S$, so as to examine its properties, such as its topological invariants, its Hodge numbers and its moduli space. This, however, is beyond the scope of this work.

Instead, we proceed by examining the properties of the surface $S$ in the limit that is dual to the stable degeneration limit in F-theory. For this approach it suffices to study the structure of generic and non-generic genus two fibers of $S$, as characterized by local models of possible genus two fibrations classified by Namikawa and Ueno in ref.~\cite{MR0369362}.

\subsection{Stable degeneration limit}
To go to the stable degeneration limit --- which relates to the heterotic large volume limit $\rho \to i\infty$ --- we would naively consider the Jacobi--Fourier expansion of the modular sections
\begin{align*}
	\psi_{2i} &= E_{2i}(\tau) + \psi_{2i,1} q_2 + \mathcal{O}(q^2_2) \ , \\
	\chi_{2j} &= \chi_{2j,1} q_2 + \mathcal{O}(q^2_2) \ ,
\end{align*}
with $q_2 = e^{2\pi i \rho}$. Consistently, we can only send $q_2$ to zero, if it is a global section. This, however, is not the case, as globally there is a non-trivial mixing with the other fiber moduli $\tau$ and $z$. 

We know from Section~\ref{sec:mspace} that the compactified moduli space of 8d heterotic theory is the blown-up weighted projective space~$\operatorname{Bl}_{h_0}\!\mathbb{P}^3_{(2,3,5,6)}$ parametrized by the homogeneous coordinates $(I_0,I_2,\psi_4, \psi_6, \chi_{10}, \chi_{12})$. Furthermore, the stable degeneration locus is identified with the exceptional divisor $H_0\equiv \pi^{-1}(h_0)$ at $\chi_{10}=\chi_{12}=0$. 

The non-geometric compactification of heterotic string is a map $\iota$ from the base $\mathbb{P}^1$ to the moduli space $\operatorname{Bl}_{h_0}\!\mathbb{P}^3_{(2,3,5,6)}$. By dimensional reasons the intersection $\iota(\mathbb{P}^1)\cap H_0$ is generically empty. Nevertheless, we can continuously send the entire image $\iota(\mathbb{P}^1)$ to the exceptional divisor $H_0$ with the help of the complex scalar parameter $\sigma$, i.e.,\footnote{This parameter is called the smoothing parameter in ref.~\cite{Louis:2011aa}}
\begin{equation}
		\chi_{10}(s) = \sigma \chi'_{10}(s), \quad \chi_{12}(s) = \sigma \chi'_{12}(s) \ .
\end{equation}
Then the stable degeneration limit becomes $\sigma\to 0$,\footnote{Formally, the homotopy $H: \mathbb{P}^1 \times [0,1] \to \operatorname{Bl}_{h_0}\!\mathbb{P}^3_{(2,3,5,6)}$ with $H(\mathbb{P}^1,1)=\iota(\mathbb{P}^1)$ and $H(\mathbb{P}^1,0)\subset H_0$ realizes the limit.}
and we have $\chi_{10}(s)\to 0$, $\chi_{12}(s)\to 0$, while the ratio $\frac{\chi_{12}(s)}{\chi_{10}(s)}$ remains finite. 

In the limit $\sigma\to 0$, we expect that the non-geometric genus two fibration becomes the geometric heterotic compactification of an elliptically fibered $K3$ surface together with an $SU(2)$ spectral cover. To arrive at this conclusions, the structure of singular genus two fibers plays an essential role.

\begin{figure}[t]
\includegraphics[width=0.32\textwidth]{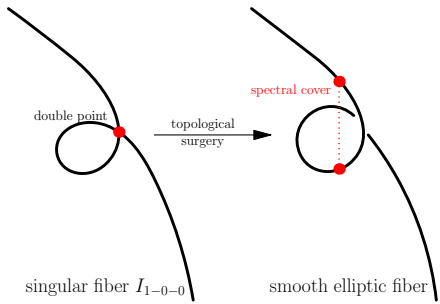}
\caption{Depicted is the topological surgery operation performed in the limit $\sigma\to0$. It maps the singular genus two fibers of type $I_{1-0-0}$ (left side) to smooth elliptic fibers with the spectral cover data of an $SU(2)$ bundle (right side).} \label{fig:I100toSmooth}
\end{figure}

Using the relations of ref.~\cite{MR1106431} among the coefficients of the genus two hyperelliptic curve \cite{MR1106431}, we determine the structure of genus two fibers in the limit $\sigma\to0$. A generic genus two fiber degenerates in the limit $\sigma\to0$ to the equation
\begin{equation} \label{eq:degg2}
  y^2 = \left(4x^3 - \frac{4}{3}\pi^4 \psi_4(s)\, x - \frac{8}{27}\pi^6 \psi_6(s)\right)
  \left( x - \gamma(s)  \right)^2 \ .
\end{equation}
Such a singular fiber is of type~$I_{1-0-0}$ in the classification of ref.~\cite{MR0369362}, which is an elliptic fiber together with a double point. Using surgery techniques we remove the neighborhood of the singular point at $x = \gamma(s)$ and glue in two disjoint smooth patches so as to arrive at a smooth elliptic fiber. In this way, we have reduced the singular fiber of arithmetic genus two to a smooth elliptic curve of genus one. Furthermore, in recording the two points $\tilde p_{\pm}$ on the elliptic fiber, where the patches have been glued in, we can deduce the spectral cover for the heterotic $SU(2)$ bundle. Given the Weierstrass function $\wp(\,\cdot\,; \tau)$ associated to the modulus $\tau$ of the elliptic curve in Weierstrass form, we find
\begin{equation} \label{eq:wp}
  \gamma(s) = \wp(\tfrac{z(s)}2;\tau(s)) \ ,
\end{equation}  
with the periods $\underline{\tau}(s)$ in the limit $\sigma\to0$. Thus, up to an overall factor of two the loci of the surgery encode the Wilson line modulus $z$ of the SU(2) bundle,\footnote{It is tempting to relate the factor of $\frac12$ in eq.~\eqref{eq:wp} to the square of the appearance of the spectral cover factor $x-\gamma(s)$ in eq.~\eqref{eq:degg2}.} and the spectral cover is given by the two points~$p_\pm = 2 \tilde p_\pm$, where $p_\pm$ arise from the zeros of the Weierstrass function $\wp(z;\tau)$. Thus, altogether the topological surgery amounts to replacing the degenerate genus two curve \eqref{eq:degg2} with the genus one Weierstrass equation
\begin{equation}
     y^2 \,=\, 4x^3 - \frac{4}{3}\pi^4 \psi_4(s)\, x - \frac{8}{27}\pi^6 \psi_6(s) \ .
\end{equation}
Furthermore, with eq.~\eqref{eq:chilimit}  the spectral cover becomes
\begin{equation} \label{eq:sc}
     0\,=\, \chi_{10}'(s)\,x-\pi^2 \chi_{12}'(s)  \ ,
\end{equation}     
which --- as required for a spectral cover of an $SU(2)$-bundle --- indeed intersects the constructed elliptic fiber in the two points $p_\pm$, which add up to zero. The described surgery process is illustrated in FIG.~\ref{fig:I100toSmooth}.

Let us now turn to the non-generic fibers in the limit $\sigma\to0$, which arise at the intersection of the image $\iota(\mathbb{P}^1)$ in $\operatorname{Bl}_{h_0}\!\mathbb{P}^3_{(2,3,5,6)}$ with (the proper transforms of) the divisors $H_1$ and $H_4$.\footnote{In Section~\ref{sec:mspace}, we have introduced the divisors $H_1$ and $H_4$ in the context of the projective space $\mathbb{P}^3_{(2,3,5,6)}$, but we use in the following the same letters for their proper transforms in the blown-up projective space $\operatorname{Bl}_{h_0}\!\mathbb{P}^3_{(2,3,5,6)}$.} 

The former intersection points correspond to singular fiber of type~$I_1$-$I_0$-$1$ of ref.~\cite{MR0369362}, which is a degenerate reducible singular fiber of two elliptic curve components with self-intersection number $-1$. In addition, one elliptic curve component has developed a nodal point. As a consequence after performing the described topological surgery on the entire surface the nodal component turns into a rational curve of self-intersection $-1$. We blow-down these rational curves to arrive at a minimal surface. Since the two elliptic components of the reducible fibers along $H_1$ intersect in the zeros of the two elliptic curves, the spectral cover points coincide after the blow-down with zero of the maintained elliptic component.

To examine the intersection points of $H_4$ with the exceptional fiber $H_0$, we determine the proper transform of the divisor $H_4$ for the blow-up \eqref{eq:blowup}. With the affine coordinate $u=\frac{I_2}{I_0}$ in the patch $I_0\ne 0$ of $\operatorname{Bl}_{h_0}\!\mathbb{P}^3_{(2,3,5,6)}$ the defining equation for the divisor $H_4$ becomes
\begin{equation}
  \frac{\chi_{35}^2}{\chi_{10}} = \chi^3_{10} (\psi_4^3-\psi_6^2)^2(27u^3 - 9 u \psi_4 - 2\psi_6) + \mathcal{O}(\chi_{10}^4) \ .
\end{equation}
Therefore, on the exceptional divisor $H_0$ --- given by $\chi_{10}=0$ --- the proper transform of $H_4$ restricts to
\begin{equation}
  \left.H_4\right|_{H_0} = (\psi_4^3-\psi_6^2)^2(27u^3 - 9 u \psi_4 - 2\psi_6) \ .
\end{equation}
In the stable degeneration limit $\sigma\to0$ of the analyzed genus two fibered surface, with eqs.~\eqref{eq:psilim} and \eqref{eq:chilimit} these two components turn into
\begin{equation}
\begin{aligned}
   D_1 &= 4 \wp(s)^3 - \frac{4\pi^4}{3}E_4(s) \wp(s) - \frac{8 \pi^6}{27} E_6(s) \ , \\
   D_2 &= (E_4(s)^3-E_6(s)^2)^2 \ .
\end{aligned}
\end{equation}
The elliptic Eisenstein functions $E_4$ and $E_6$ are now holomorphic sections of $\mathcal{O}(8)$ and $\mathcal{O}(12)$, respectively, while $\wp$ is a meromorphic section of $\mathcal{O}(4)$ over the base $\mathbb{P}^1$. 

At the points $D_1=0$, we again obtain singular genus two fibers of the type $I_{1-0-0}$, which turn into smooth elliptic fibers via the surgery. At such points the Wilson line modulus $z$ becomes a half elliptic period,\footnote{This is because $D_1$ is identified with $(d \wp(s)/dz(s))^2$.}, which shows the appearance of a non-generic $\mathbb{Z}_2$ fiber symmetry.

At the points with $D_2=0$ we find singular genus two fibers of type~$I_{1-1-0}$ in the classification of ref.~\cite{MR0369362}. These are genus two fibers with two ($\mathbb{Z}_2$ symmetric) nodal points, where the two nodal points in the genus two fibers relate to the square in the component $D_2$. The topological surgery removes one nodal point, and we are left with a singular elliptic fiber of Kodaira type~$I_0$.

Altogether --- after implementing the topological surgery and performing the blow-down to a minimal surface --- we arrive at a smooth elliptically fibered surface over $\mathbb{P}^1$ with the discriminant $\Delta = E_4(s)^3 - E_6(s)^2$, which is a section of $\mathcal{O}(24)$. As a consequence the constructed surface has $24$ singular elliptic fibers of Kodaira type~$I_0$, each of which descends from a degenerate genus two fiber over the intersection of $\iota(\mathbb{P}^1)$ and the divisor $H_4$. However, this is nothing else but an elliptically fibered $K3$ surface of the Weierstrass form
\begin{equation}
  0 = -y^2 + 4 x^3 - g_2(s) x - g_3 (s) \ ,
\end{equation}
where $g_2(s)=\frac{4\pi^4}3E_4(s)$ and $g_3(s)=\frac{8\pi^6}{27}E_6(s)$ are now sections of $\mathcal{O}(8)$ and $\mathcal{O}(12)$. Furthermore, it comes with the spectral cover~\eqref{eq:sc} of a $SU(2)$ bundle. Thus, the stable degeneration limit in F-theory indeed realizes the limit of the non-geometric heterotic compactification to the geometric semi-classical large volume heterotic compactification on an elliptic $K3$ surface together with an $SU(2)$ bundle.

\subsection{Heterotic Sen limit}
As the non-geometric F-theory--heterotic duality for 4d and 6d theories gives us insight in the hypermultiplet moduli space beyond the semi-classical supergravity Kaluza--Klein reduction of the heterotic string, we can use this correspondence to study quantum effects in the hypermultiplet sector of the heterotic string. For concreteness we focus now on the 4d theories. 

In ref.~\cite{Alexandrov:2014jua} Alexandrov, Louis, Pioline and Valandro study the heterotic hypermultiplet sector in a similar context. They utilize the duality between the heterotic string on $K3\times T^2$ and the type~IIB string on the mirror $Y$ of the dual elliptically fibered Calabi--Yau threefold $X$. They work in a limit, where the volume of $K3$ is very large to suppress $\alpha'$ corrections. Furthermore, they demand that the base $\mathbb{P}^1$ of the elliptically fibered $K3$ is much larger than the elliptic fiber. In the dual type~IIB theory this renders both the 4d string coupling $g_s^{\textrm{4d}}$ and the 10d string coupling $g_s^{\textrm{10d}}$ small. As a consequence all types of string corrections become negligible. As a result, using the classical c-map metric of the hypermultiplet sector in the type IIB theory \cite{Cecotti:1988qn,Ferrara:1989ik}, they find that the metric of the heterotic hypermultiplet moduli space exhibits a two-staged fibrational structure:
\begin{equation} \label{eq:2fib}
   \vcenter{\halign{\hfil$#$&\hfil\,$#$\,\hfil&\hfil$#$\hfil\cr
      \mathcal{M}_B(g,F) & \longrightarrow & \mathcal{M}_H \cr
      &&\downarrow \cr
      \mathcal{M}_F(g) & \longrightarrow & \mathcal{M}_{g,F}\cr
      &&\downarrow \cr
      && \mathcal{M}_g \cr}}
\end{equation}
Here the bundle moduli space~$\mathcal{M}_F(g)$ is fibered over the moduli space $\mathcal{M}_g$ of the metric of the $K3$ surface due to the anti-self-dual field strength $\star_g F = - F$ given in terms of the metric-dependent Hodge star $\star_g$ of the $K3$ surface. Then the hypermultiplet moduli space $\mathcal{M}_H$ is completed by the B-field moduli space $\mathcal{M}_B(g,F)$, which in turn is fibered over the moduli space $\mathcal{M}_{g,F}$. The B-field is governed by the supergravity equation of motion $H = dB + \frac{1}{4} (\omega_G - \omega_L)$ and $d \star H = 0$ with the gravitational and gauge Chern--Simons terms $\omega_G, \omega_L$, which depend on the metric and the gauge connection.

By construction the two-staged fibration arises from the supergravity approximation of the heterotic Kalaza--Klein reduction. As a consequence, deep inside the hypermultiplet moduli space --- where in particular $\alpha'$-corrections are sizeable --- we expect this structure to break down. With the help of the non-geometric heterotic compactification, we present an argument here that this fibrational structure indeed disappears in the interior of the heterotic hypermultiplet moduli space.

Let us suppose that in the Weierstrass model of the F-theory compactification \eqref{equ:6dWeierstrassModel} is determined by the sections 
\begin{equation}
\begin{aligned}
  \psi_4(s) &= \alpha_4 h(s)^2 \ , & \psi_6(s) &= \alpha_6 h(s)^3 \ ,\\
  \chi_{10}(s) &= \alpha_{10} h(s)^5 \ , & \chi_{12}(s) &= \alpha_{12} h(s)^6 \ ,
\end{aligned}
\end{equation}
where $h(s)$ is a section of the line bundle $\mathcal{O}(4)$ over the base $\mathbb{P}^1$ while $\alpha_4,\alpha_6,\alpha_{10},\alpha_{11}$ are constants. In the dual non-geometric heterotic string, any generic point on the $\mathbb{P}^1$ base describes the same genus two curve in the (compactified) moduli space $\overline{\mathcal{A}}_2$ of genus two curves. As a consequence the genus two period $\underline{\tau}$ is constant over the entire base. Nevertheless, there are still degenerate genus two fibers, appearing at the zero loci of the section $h(s)$. We call this limit the \emph{heterotic Sen limit}, as this picture realizes the analog for genus two fibrations to the conventional Sen limit for elliptic fibrations in the context of F-theory \cite{Sen:1997gv}.

For our purpose, in the heterotic Sen limit the three constant periods $\underline{\tau}$ of the genus two fibers are regarded as three moduli fields in the heterotic theory. Subsequently, we identify the complex structure moduli field $\tau$, the complexified K\"ahler moduli field $\rho = B + i J$, and the gauge bundle moduli field $z$ with a subset of two scalar degrees of freedom of three hypermultiplets \cite{Louis:2011aa}. In the semi-classical limit these three fields are distributed over the entire two-staged fibrational structure \eqref{eq:2fib}. Nevertheless, in the heterotic Sen limit, this property clearly breaks down, because neither is the metric for $z$  fibered over $\tau$ or $J$, nor is the metric of the $B$-field fibered over the remaining fields. Instead, there is a $\mathbb{Z}_2$-symmetry exchanging $\tau$ and $\rho=B+ iJ$. This can be seen formaly, as the $SO(2,3)$-isometric metric on the sub-moduli space $\mathcal{M}_{\textrm{sub}}$ parametrized by $\tau, \rho$, and $z$ is the same as the metric on the K\"{a}hler symmetric space $\mathcal{D}_{2,3}$, which is explicitly spelt out in ref.~\cite{Alexandrov:2012pr}. For the pairing $\eta$ of the lattice~$L^{2,3}$, represented by the matrix 
\begin{equation}
		\eta_{AB}= \begin{pmatrix}
		0 & 1 & 0 & 0 & 0\\
		1 & 0 & 0 & 0 & 0\\
		0 & 0 & 0 & 1 & 0\\
		0 & 0 & 1 & 0 & 0\\
		0 & 0 & 0 & 0 & -2\\
		\end{pmatrix} \ ,
\end{equation}
we parametrize a null vector $X$ in terms of the three moduli $\tau,\rho,z$ as \cite{MR2935386,Alexandrov:2012pr}
\begin{equation}
		X^A = \left( 1, z^2 - \tau \rho, \tau, \rho, z \right) \ .
\end{equation}
Then one finds that $\eta_{AB}X^AX^B = 0$ and $\eta_{AB}X^A\bar{X}^B >0$, and the metric on $\mathcal{D}_{2,3}$ can be written as \cite{Alexandrov:2012pr}
\begin{equation}\label{equ:D23metric}
		ds^2_{\mathcal{M}_{\textrm{sub}}} =  \mathcal{K}_{i\bar{\jmath}} dt^i d\bar{t}^{\bar{\jmath}} \ , \quad t^i = (\tau,\rho,z) \ ,
\end{equation}
with $\mathcal{K}_{i\bar{\jmath}}=\partial_i\partial_{\bar{\jmath}} \mathcal{K}$ and the K\"{a}hler potential 
\begin{equation}
		\mathcal{K} = -\frac14\log \left(\eta_{AB} X^A \bar{X}^B \right) \ .
\end{equation}
The obtained K\"ahler metric \eqref{equ:D23metric} with $SO(2,3)$ isometry --- representing a subspace of the quanternionic hyper-K\"ahler metric of the hypermultiplet moduli space --- clearly does not fit in the two-staged fibration \eqref{eq:2fib}.

\section{Conclusions} \label{sec:con}
Using the non-geometric duality correspondence proposed by Clingher--Doran and Malmendier--Morrison \cite{MR2824841,MR2935386,Malmendier:2014uka}, we studied F-theory--heterotic duality beyond the semi-stable degeneration limit and the semi-classical heterotic limit in eight and lower space-time dimensions. By geometric means, we analyzed in detail the limit from the non-geometric quantum phase to the semi-classical heterotic phase. This allowed us to argue that in six and four space-time dimensions, which arose from adiabatically fibering the eight dimensional duality over a common base, the non-geometric heterotic compactifications captured $\alpha'$-corrections to the semi-classical large fiber compactification. Furthermore, we shed light on the puzzling phenomenon discovered in ref.~\cite{Malmendier:2014uka} that even though the heterotic string moduli space continuously interpolated between the non-geometric quantum phase and the semi-classical large fiber phase that the dual F-theory description remained geometric over the entire dual moduli space. 

Analyzing the heterotic theory in four dimensions in the heterotic Sen limit, we observed that $\alpha'$-corrections modified the semi-classical two-staged fibrational structure derived by Alexandrov, Louis, Pioline and Valandro \cite{Alexandrov:2014jua}. While such alterations to the hypermultiplet sector are expected on general grounds \cite{Alexandrov:2013yva}, we believe that the analysis of non-geometric heterotic compactifications in the quantum regime --- in particular at special loci in the moduli space such as the heterotic Sen limit --- may shed light on conceptual questions concerning the quantum hypermultiplet moduli space.

To make further progress in analyzing the discussed class of non-geometric heterotic string compactifications in four and six dimenions, we need a better understanding of the relevant genus-two fibered surfaces. In this work, we mainly analyzed the local structure of such genus two fibered surfaces by studying the structure of their singular fibers, using the classification of Namikawa and Ueno \cite{MR0369362}. It would be interesting to study the global features of the relevant genus-two fibered surfaces --- for instance by comparing to the general properties of genus two fibered surfaces~\cite{MR2316980} --- and to give more detailed physical interpretations of the geometric structures of such surfaces.

While the studied F-theory--heterotic quantum duality is based on the rather special class of F-theory--heterotic models with a single Wilson line modulus, it provides for an explicit and detailed toy model for non-geometric string compactifications in general. The technique --- to combine the compactification space with the spectral cover data by surgery operations --- may also open up a new method to arrive at more general non-geometric heterotic string theories, which describe quantum corrected heterotic strings beyond a single non-trivial Wilson line modulus.

Although our discussion starts from the F-theory--heterotic duality, the general philosophy to use local duality transformations in order to arrive at non-trivial non-geometric global string compactifications \cite{Hull:2004in}, is applied here as well. The possibility to explicitly describe the transition to a conventional semi-classical heterotic compactification in terms of topological surgery makes our non-geometric heterotic model appealing and calculable. It would be interesting to see if such non-geometric models also relate to recent proposals on non-geometric string compactifications \cite{Blumenhagen:2011ph,Blumenhagen:2012nt,Hohm:2013bwa,Bakas:2013jwa}.
\bigskip
\section*{Acknowledgements}
We would like to thank 
Per Berglund,
Fabrizio Catanese,
Chuck Doran,
Daniel Huybrechts,
Albrecht Klemm,
Andreas Malmendier, 
Stephan Stieberger,
and
Thomas Wotschke
for interesting discussions and useful correspondences.
J.G. is supported by the BCGS program, and H.J. is supported by the DFG grant KL 2271/1-1.


%
\end{document}